# Denoising Shack Hartmann Sensor spot pattern using Zernike Reconstructor

Akondi Vyas[1], M B Roopashree[2], B Raghavendra Prasad[3]

[1]Research Scholar, Indian Institute of Astrophysics, Bangalore-34
[2]Research Scholar, Indian Institute of Astrophysics, Bangalore-34
[3]Professor, Indian Institute of Astrophysics, Bangalore-34
vyas@iiap.res.in, roopashree@iiap.res.in, brp@iiap.res.in.

**ABSTRACT**

*Shack Hartmann Sensor (SHS) is inflicted with significant background noise that deteriorates the wave-front reconstruction accuracy. In this paper, a simple method to remove the back ground noise with the use of Zernike polynomials is suggested. The images corresponding to individual array points of the SHS at the detector, placed at the focal plane are independently reconstructed using Zernike polynomials by the calculation of Zernike moments. Appropriate thresholding is applied on the images. It is shown with computational experiments that using Zernike Reconstructor along with usual thresholding improves the centroiding accuracy when compared to direct thresholding. A study was performed at different Signal to Noise ratio by changing the number of Zernike orders used for reconstruction. The analysis helps us in setting upper and lower bounds in the application of this denoising procedure.*

## 1. INTRODUCTION

There are a wide range of applications where wave-front sensing is essential like adaptive optics, lens testing and ophthalmology [1]. A Shack Hartmann Sensor (SHS) measures the shape of the wave-front incident on it [2]. It is made up of lenses arranged in a two dimensional array that directs the light towards the focal plane. A detector placed at the focal plane of the lenslet array records the position of the spots. The spots by and large maintain a Gaussian structure. The signatures of the distortions in the wave-fronts are seen by the sensor as local shifts in the focal spots depending on the local slopes across individual apertures [3]. The wave-front sensors generally operate in real-time and the position of the fluctuating spot is estimated at a frequency equal to the rate of fluctuations using centroiding techniques [4]. The shape of the wave-front is then evaluated from the measured shifts of the spots [5-8].

The problem of estimating the centroid of an incoherently imaged point with a detector array is analyzed in the literature [9]. The effects of this on a wave-front reconstruction formed by a Shack–Hartmann sensor are described. Generally, the wave-fronts falling on the sensor carry along with them additive noise which cannot be distinguished most times from the actual signal. There are many methods of minimizing the effects of noise like thresholding, noise filtering by linear, median and adaptive filters. In applications where the Signal to Noise Ratio (SNR) is small, noise minimization becomes important. At low SNR, some of the above mention methods of noise minimization fail. They even do not take the advantage of the fact that in this case the spot pattern is generally Gaussian.

Zernike polynomials are a set of continuous orthogonal circular polynomials defined over the unit disk. Since they form a complete set of orthogonal polynomials, any 2D function can be represented as a proper linear combination of this basis set [10]. Zernike polynomials are used in many applications such as pattern recognition, image representation, aberration production and wave-front sensing [11-12]. Zernike polynomials are known for their ease of production and representation of Seidel aberrations using lower order Zernike polynomials. There are many recursive algorithms for easy computation of Zernike moments of two dimensional image functions [12]. In this paper, calculation of the Zernike moments is done using a fast and accurate method implemented by Hosny that minimizes the geometric errors by a proper image mapping and removes approximation errors by the exact calculation of Zernike moments [13]. Noise is generally a higher order feature compared to the signal. The reconstruction of the images using lower order Zernike moments will minimize the noise amplitude compared to the signal. This method may fail if the spatial extent of noise becomes significantly comparable to the signal length scales, in this case, the spot size. By applying a suitable noise threshold over the images, it is also possible to eliminate the effects of background noise. In the case of the Shack Hartmann spots, the signal features are similar to noise features for small spots and hence more care has to be taken so that the signal strength is not reduced. This is the major reason why regular noise removal algorithms fail in this case.

The performance of the proposed method is a function of the SNR, the number of Zernike moments used for image reconstruction and threshold. A detailed analysis of the centroiding accuracy to the sensitivity of SNR, optimum number of Zernike moments and threshold limit is presented. Optimization of these parameters will allow us to place limits over the performance of the sensor in the presence of noise. Taking the advantage of the fact that the spot maintains a Gaussian shape, it is shown using computational experiments that the noise can be efficiently removed even at low signal levels.

## 2. BACKGROUND

Zernike polynomials are well known basis for image processing and image representation applications. The finer details of the images are represented using higher orders of Zernike moments and the broad features need computation of lesser number of moments.

The Shack Hartmann spot pattern superimposed by a uniform background noise was simulated. These simulated images were used for the statistical analysis of the proposed method. The spot pattern images were reconstructed using Zernike polynomials.

### 2.1 Simulation of the spot pattern

The Simulation of the spot pattern at the focal plane of SHS was performed in two steps as detailed below:
- A 2D Gaussian intensity pattern was simulated with an image size of 64×64 square pixels. The centre of the spot was positioned at a known position on the image. The shifts were measured with respect to the image centre. The assumption of a Gaussian image spot is more appropriate in the case of larger spots since they satisfy the minimum distance condition to represent a Gaussian spot better. The simulation has control over the spot size, SNR, the shift of the spot and image size. Throughout the paper, the spot size was maintained to be 6 pixels across the diameter of the spot. Spot size is defined as the distance over which the intensity falls off to 1/e value of the maximum intensity.
- A 2D spatially uniform noise was then superimposed on the simulated spot whose spatial intensity distribution function followed Gaussian statistics.

Sample spot pattern images with different SNR are shown in Fig. 1.

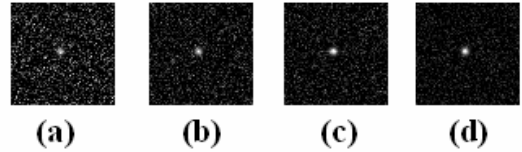

Fig. 1. Spot pattern images with SNR (a) 0.3 (b) 0.5 (c) 0.7 and (d) 0.9

### 2.2 Proposed Method-Zernike Reconstructor (Z.R) + Thresholding (Th)

The proposed noise removal method is a two step process involving the reconstruction of images using Zernike polynomials and the application of classical image thresholding algorithm. The spot pattern image with SNR=0.3 reconstructed using Zernike moments is shown in Fig. 2.

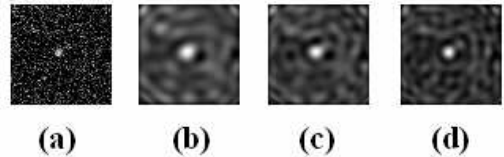

Fig. 2. Images reconstructed using Zernike Polynomials (a) original image (b) Reconstructed using 25 orders (c) 30 orders (d) 35 orders

During the thresholding process, individual pixels are marked as target pixels if their pixel value is greater than the threshold and pixels with pixel values below the threshold are forced to take the lowest pixel value. Applying thresholding (80% of the peak value) to the noise image directly and the reconstructed images shown in Fig. 2(b), 2(c), 2(d) have a different effect and are shown in Fig. 3.

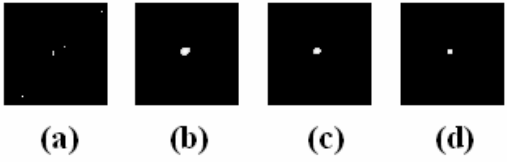

Fig. 3. Thresholding (a) original image (b) 25 orders (c) 30 orders (d) 35orders

## 2.3 Centroiding Algorithms

The centroiding methods used in this paper include normalized Centre of Gravity (CoG) and Iteratively Weighted Centre of Gravity (IWCoG). The CoG method uses the averaging formula, which is the ratio of sum of products of position coordinate and intensity at that point to the total intensity. Weighted CoG method uses additional information that the spot has a Gaussian spread and weights the intensity function with a Gaussian intensity distribution, effectively fitting a Gaussian to the spot. In IWCoG method, the position of the Gaussian centre and the spread are iteratively corrected to go closer to the actual position of the centroid [4].

These techniques have their advantages and disadvantages. CoG method has advantage over other techniques in the absence of noise. In the presence of background noise the performance of CoG method is degraded. If the background noise is uniform and the number of pixels under observation (image size) is large then statistically, the centroid of the noise will be closer to the image centre and not close to the actual centroid at low SNR. The IWCoG method has an advantage that it can detect the centroid position more accurately even in the presence of noise, but at the cost of increased computational time and iteration convergence problems. If the shape of the spot is not retained as a near Gaussian due to the addition of background noise, IWCoG fails to accurately locate the centroid, in which case noise minimization becomes critical.

## 2.4 Average Centroid Estimation Error

The performance of the centroiding algorithms was analysed using the percentage centroid estimation error (CEE) defined as shown below:

$$CEE = \frac{\sqrt{(x_c - x_c^*)^2 + (y_c - y_c^*)^2}}{S} \times 100 \quad (1)$$

where $(x_c, y_c)$ represents the position of the actual centroid (known since the spot position is controlled by the simulation), and $(x_c^*, y_c^*)$ is the centroid position estimated using the algorithms. S is the amplitude of shift of the spot from the image centre (also known from simulation). Since the added noise is a statistical quantity, average centroid estimation is calculated which is an average of many realizations of the centroid estimation error for a particular case. In the subsequent sections, average centroid estimation error is the mean of 100 realizations of the centroid estimation error.

## 2.5 Gaussian Spot

In this case it is possible to take the advantage of the fact that the spot maintains a Gaussian shape. Fig. 4 shows the case of partial noise removal.

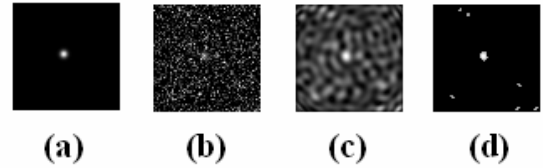

Fig. 4. Partially noise removal (a) Gaussian spot (b) Noise imposed Gaussian spot, SNR=0.2 (c) Spot pattern reconstructed using 44 orders of Zernike polynomials (d) 60% Thresholding.

The minor features in Fig. 4(d) are a resultant of the significant features in the reconstructed spot, Fig. 4(c). Although these minor features are significantly large, they do not bear a Gaussian shape. Pattern recognition algorithms can be used to eliminate these features. The next question arises as to how to recognize whether the images are partially or fully noise free. This can be done by measuring their correlation with a standard Gaussian spot image. If the images contain these minor features, the value of the correlation coefficient is low ($< 0.7$).

## 3. SIMULATION RESULTS

### 3.1 Signal to Noise Ratio

The performance of the centroiding algorithms depends strongly on the SNR. A comparison of CoG and IWCoG algorithms at different SNR is shown in Fig. 5.

The performance of the Zernike reconstructor based noise removal algorithm method is a function of the signal to noise ratio when applied to different centroiding algorithms. The comparison of the performance of the CoG algorithm before and after the introduced noise removal algorithm is shown in Fig. 6.

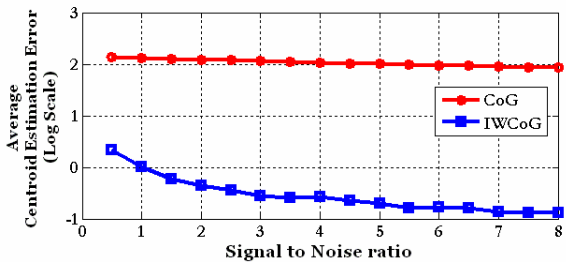

Fig. 5. Comparison of the performance of the centroiding algorithms

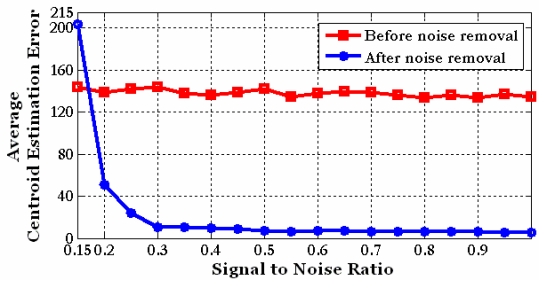

Fig. 6. Comparison of the performance of CoG method before and after noise removal algorithm

Fig. 6 suggests that the application of noise removal algorithm has very significant effect on the CoG algorithms at low SNR which otherwise is an under performer. Above a SNR of 0.3, the centroid estimation error reduces below 10%. The same effect is not seen in the case of IWCoG. The application of the noise removal algorithm is shown in Fig. 7. As it is, the IWCoG method has little CEE at SNR > 0.5 as shown in Fig. 5. The extent of improvement in the performance of the algorithm at SNR = 0.25 is greater than 5%.

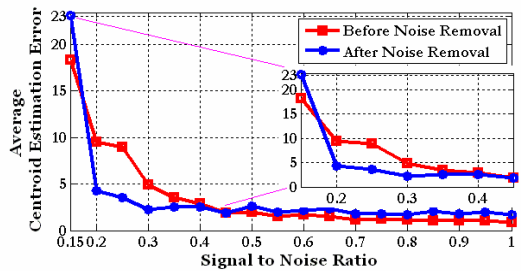

Fig. 7. Comparison of the performance of IWCoG method before and after noise removal algorithm

### 3.2 Shift in spot

The CEE reduces in the case of higher shift in the spots in the case of CoG and IWCoG in general without applying the noise removal algorithm. The behavior of the CoG algorithm with ZR + Th for shifts of 0.5, 1, 1.5 pixels at different SNR also follows a similar trend below a SNR of 0.4. The bigger spot size has lesser error as shown in Fig. 8. There is an interesting phenomena observed repeatedly at SNR = 0.4, where the curves with 0.5 and 1.0 pixel shift cross each other.

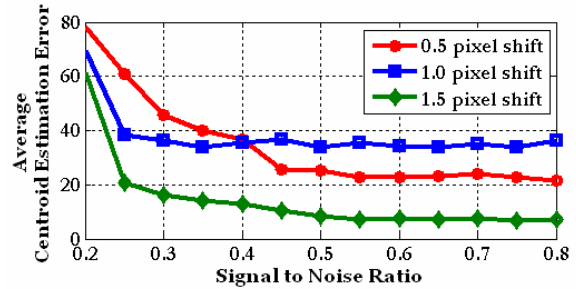

Fig. 8. CoG algorithm performance (with ZR+Th) at different shift amplitudes

### 3.3 Zernike Orders

The effect of using different number of Zernike moments for reconstruction is shown in Fig. 2 and Fig. 3. Increasing the number of Zernike moments for image reconstruction makes the finer features of the image more prominent. And hence the noise which is a finer feature in our case stands out. Making the finer features more prominent puts a lower limit on the threshold for complete noise removal, but this higher thresholding may lead to signal loss too. Hence it is suggested to use less number of Zernike orders for reconstruction and a hence lower threshold for total noise removal.

### 3.3 Thresholding Limits

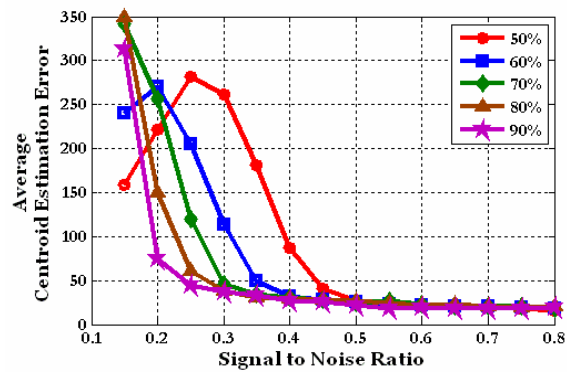

Fig. 9. Choice of threshold below SNR=0.5 maybe critical

Thresholding depends on the signal to noise ratio. A lower limit on thresholding is dependent on noise and the upper limit on thresholding is dependent on the signal strength. The dependence of the centroiding accuracy on the choice of threshold is shown in Fig. 9. The choice of threshold has no effect for SNR > 0.5.

## 4. CONCLUSIONS

The Zernike reconstructor along with suitable thresholding on images of the Shack Hartmann spot pattern can become a very effective tool for close to complete noise removal. This method can also be applied to cases where the signal to noise ratio is small and the spatial extent of the noise is much smaller compared to the signal. To avoid IWCoG that leads to convergence problems, the option of using CoG algorithm along with Zernike reconstructor based noise removal algorithm may be considered in wave-front sensing applications. It is shown that the accuracy of centroiding improves nearly 20 times while using CoG algorithm in the presence of noise with SNR less than 1. It is shown through computational experiments that even at low amplitudes of shifts this algorithm performs a good job. The limits on the threshold value is analyzed and presented.